\documentclass[12pt]{article}

\usepackage{graphicx}
\usepackage{amsfonts}
\usepackage{amssymb}
\usepackage{amsmath}
\usepackage[colorlinks,urlcolor=blue]{hyperref}

\begin{document}

\title{\bf Quaternions in Hamiltonian dynamics of a rigid body -- Part II.\\
Relation of canonical Poisson and Lie-Poisson structures}
\providecommand{\keywords}[1]{\textbf{\textit{Keywords: }} #1}

\author{
  Stanislav S. Zub\\
  Faculty of Cybernetics,\\
  Taras Shevchenko National University of Kyiv,\\
  Glushkov boul., 2, corps 6.,\\
  Kyiv, Ukraine 03680\\
  \texttt{stah@univ.kiev.ua}\\
  \\
  Sergiy I. Zub\\
  Institute of Metrology,\\
  Mironositskaya st., 42,\\
  Kharkiv, Ukraine 61002\\
  \texttt{sergii.zub@gmail.com}}

\maketitle

\newpage
\begin{abstract}
It was proposed the Lie group such that symplectic structure of orbits
of co-adjoint representation of the group is revealed symplectic structure
of a rigid body dynamics in quaternion variables. It is shown
that Poisson brackets of corresponding Lie-Poisson structure
coincide with canonical Poisson brackets on cotangent bundle of group unit quaternions.

\keywords{quaternion, symplectic structure, Poisson structure, Lie-Poisson brackets, Liouville form, Kirillov-Kostant-Souriau form, SE(3), SO(3) group}.

\end{abstract}

\newpage
\tableofcontents

%
\newcommand{\s}[1]{\ensuremath{\boldsymbol{\sigma}_{#1}}}
\newcommand{\bsym}[1]{\ensuremath{\boldsymbol{#1}}}
\newcommand{\w}{\ensuremath{\boldsymbol{\wedge}}}
\newcommand{\lc}{\ensuremath{\boldsymbol{\rfloor}}}
\newcommand{\rc}{\ensuremath{\boldsymbol{\lfloor}}}

\thispagestyle{empty}

\newpage
\section{ Introduction }
\label{Intro}

\bigskip
A great number of publications are devoted to application of quaternions in rigid body mechanics.
The majority of them belong to the kinematics of a rigid body that is describe the orientation of a rigid body in space by using quaternion parameters \cite{Whittaker:NYork44}, but only a few works are devoted to application of quaternions in dynamics.

Lagrangian description of a rigid body dynamics in quaternions is given by Kozlov \cite{Kozlov:Igevsk95}
the work based on classical approach of Poincare. Borisov and Mamaev proposed expressions for Poisson brackets
between quaternion parameters and components of intrinsic angular momentum of a body \cite{BM:NonLinePS97,BM:PS99,BM:RBD01}.
From deep relations of quaternion algebra and $SO(3)$ and $SO(4)$ groups one can considered these variables as generators
of Lie-Poisson structure associated with $SO(4)$ group.

In our work \cite{Zub_Quat1} quaternion parameters are regarded as dynamic variables in Hamiltonian dynamics of a rigid body.
Thus Poisson brackets that were obtained in works \cite{BM:NonLinePS97,BM:PS99,BM:RBD01}
are simply a subset of Lie algebra of Poisson brackets for dynamical variables of a rigid body.
There is only one specifics of such consideration that well known two-valuedness of these variables.

In work \cite{Zub_Quat2} it is shown that Poisson brackets of quaternion variables corresponds to canonical symplectic structure
on cotangent bundle group of unit quaternions.
Thus in one mathematical model was combined two mathematical structures proposed by Hamilton, i.e. quaternion algebra and Hamiltonian formalism.

The work \cite{Zub_Quat3} discusses geometric and algebraic aspects of Hamiltonian formalism
for a rigid body in quaternion variables and also its application to description of {\it asymmetric}
small magnetic body (or dipole) dynamics in an external constant magnetic field.

In works \cite{Zub_Quat1,Zub_Quat2,Zub_Quat3} quaternionic Poisson brackets are regarded as a canonical
that are correspond to Liouville form on $T^*SO(3)$, $T^*SE(3)$ and $T^*S^3$.
On other hand it is obvious that these brackets are linear in respect to variables,
i.e. they have form of Lie-Poisson brackets that is leading idea of the pioneering works \cite{BM:NonLinePS97,BM:PS99,BM:RBD01}. The question is what are relations between these two aspects.

In this work we proposed Lie group such that symplectic structure of orbits
of co-adjoint representation of the group is revealed symplectic structure
of a rigid body dynamics in quaternion variables.
Here we specify on a relation of symplectic structure of Kirillov-Kostant-Souriau
on orbits of coadjoint representation of this group and canonical symplectic structure
on cotangent bundles of spheres in 4-dimensional space of quaternions with standard metric.

This study of orbits is similar on consideration of orbits of co-adjoint representation of group $SE(3)$
(see \cite{Marsden:IMS94,MarMisOrtPerRat07,ZubHDLevOrb}).
Since as the Lie-Poisson structure that we will study below has only one Casimir function so there are two,
but not three orbits types.
Then orbits of 1-st type are completely analogous to those that in work \cite{Marsden:IMS94}.
The orbits of 2-nd type have an other dimensionality but in many ways similar to orbits of 2-nd type in the work.
As for 3-rd type orbits are characterized of a magnetic member that is added to Kirillov-Kostant-Souriau form
but such orbits does not arise in this group.

\newpage
\section{ The Lie algebra associated with quaternionic Poisson brackets }
\label{AlgebraLie}

\bigskip
As is known position of a rigid body that rotating around the center of mass can be described as a quaternion $q=q^0 e_0+q^1 e_1+q^2 e_2+q^3 e_3$ (see \cite{BM:PS99,BM:RBD01,Zub_Quat1}).
Intrinsic angular momentum of a rigid body also can be described as a quaternion, precisely, as a pure quaternion $\bsym{\mu}=\mu^1 e_1+\mu^2 e_2+\mu^3 e_3$.

Their components $q^{\alpha}$, $\alpha=0,\dots,3$ and $\mu^i$, $i=1,\dots,3$
can be seen as a subset of dynamical variables in Hamiltonian formalism for a rigid body (see Part I or \cite{Zub_Quat1})
that are completely describe the body dynamics.

Then in a reference system associated with a body (see (10), page 39 in Part I, \cite{Zub_Quat2}
the representation of left trivialization) we have the following base Poisson brackets.
\[
\begin{cases}
   \{\mu_i, \mu_j\} = - 2\varepsilon_{ijk}\mu_k; \\
   \{\mu_i, q_0\} =  q_i; \\
   \{\mu_i, q_j\} = - q^0\delta_{ij} - q^k\varepsilon_{ijk}; \\
   \{q_\mu, q_\nu\} = 0
\end{cases}
\leqno(1)\]

{\bf Remark 1}. In contrast to Part I $\mu^i$ stand for angular momentum of a rigid body in {\it left} trivialization.

These relations can be written in coordinateless form with using quaternion algebra \cite{Zub_Quat1,Zub_Quat2}, so
\[
\begin{cases}
   \{\bsym{\mu}, \langle\bsym{\mu},\bsym{\xi}\rangle\}
   = ad^*_{\bsym{\xi}}[\bsym{\mu}] = 2\bsym{\mu}\times\bsym{\xi}; \\
   \{q,\langle\bsym{\mu},\bsym{\xi}\rangle\} =  q\bsym{\xi} = L_q\bsym{\xi}; \\
   \{\langle q, a\rangle,\langle q, b\rangle\} = 0
\end{cases}
\leqno(1a)\]
where $\bsym{\xi}$ -- a fixed pure quaternion,
$a, b$ -- the fixed quaternions.

Despite the fact that in works these Poisson brackets were received as a canonical,
i.e. such that correspond to canonical symplectic structure on $T^*S^3$($T^*SO(3)$)
that are follow from Liouville form. They have form of the Lie-Poisson brackets,
i.e. they are linear in variables $q,\bsym{\mu}$.

For the first time brackets (1) were obtained in work of Borisov and Mamaev while investigating of application of Lie algebra of group $E(4)$ in mechanics problems \cite{BM:NonLinePS97}.

Logically to set a problem of finding the Lie group that direct connected with Lie-Poisson structure determined from formulas (1).
In particular dimensionality of this group should be 7 instead of 10 as it appears in work \cite{BM:NonLinePS97}.

Let's consider the Lie algebra of dimensionality 7 with basis elements:\\
$\bsym{\varepsilon}_i$, $i=1,\dots,3$ and $e_\alpha$, $\alpha=0,\dots,3$

If we assume that Poisson bracket with sign "$-$" corresponds to left trivialization \cite[(13.1.1), p. 426]{MarRat98}
then we obtain the following basic relations for the Lie algebra $\bsym{g}=\mathbb{H}_0\oplus\mathbb{H}$ 
\[ \begin{cases}
   [\bsym{\varepsilon}_i, \bsym{\varepsilon}_j] = 2\varepsilon_{ijk}\bsym{\varepsilon}_k; \\
   [\bsym{\varepsilon}_i, e_0] =-e_i; \\
   [\bsym{\varepsilon}_i, e_j] = \delta_{ij} e^0 + \varepsilon_{ijk}e^k; \\
   [e_\mu, e_\nu] = 0
\end{cases}
\leqno(2)\]

Using quaternion algebra as in (1a) we will obtain relations equivalent to~(2)
\[
\begin{cases}
   [\bsym{\xi}, \bsym{\eta}] = 2\bsym{\xi}\times\bsym{\eta}; \\
   [\bsym{\xi}, \nu] =  -\nu\bsym{\xi}^\flat; \\
   [\nu, \bsym{\xi}] =  \nu\bsym{\xi}^\flat; \\
   [\nu, \nu'] =  0;
\end{cases}
\leqno(3)\]
where "${}^\flat$" --- a linear isomorphism $\bsym{\varepsilon}_i \rightarrow e_i$.

Same relations can be written more compactly in form
\[ [\bsym{\xi} + \nu, \bsym{\xi}' + \nu'] = 2\bsym{\xi}\times\bsym{\xi}'
   + \nu\bsym{\xi}'^\flat - \nu'\bsym{\xi}^\flat
\leqno(3a)\]

\newpage
\section{ Lie group associated with quaternionic Poisson brackets }
\label{GroupLie}

\bigskip
Introduce the notation:

$\mathbb{H}$ --- a quaternion algebra;

$\mathbb{H}_0$ --- a linear space of pure quaternions
that is also the Lie algebra with respect to commutator we will write with bold symbol;

$\mathbb{H}_1$ --- a subset of unit quaternions, i.e. quaternions of unit length that is the Lie group with Lie algebra $\mathbb{H}_0$;

$S^3 = \mathbb{H}_1$ --- a unit sphere in $\mathbb{H}$ that regarded as a 4-dimensional Euclidean space.

\bigskip
Consider the group $G$ of a quaternionic $2\times 2$-matrices of the form
\[ G = \left\{ g =
\begin{bmatrix}
          1 & q \\
          0 & s
   \end{bmatrix}:
   \quad q\in\mathbb{H},\quad s\in S^3
   \right\}
\leqno(1)\]

Then
\[  g g' =
   \begin{bmatrix}
          1 & q \\
          0 & s
   \end{bmatrix}
   \begin{bmatrix}
          1 & q' \\
          0 & s'
   \end{bmatrix} =
   \begin{bmatrix}
          1 & q' + q s' \\
          0 & s s'
   \end{bmatrix}
\leqno(2)\]

For inverse element
\[  g^{-1} =
   \begin{bmatrix}
          1 & q \\
          0 & s
   \end{bmatrix}^{-1} =
   \begin{bmatrix}
          1 & -q s^{-1} \\
          0 & s^{-1}
   \end{bmatrix}
\leqno(3)\]

Lie group $G$ is semidirect product
\( G = S\circledS H\), where group $H$ ---
a vector space $\mathbb{H}$
that is regarded as an Abelian group with respect to addition and $S=S^3$.

Hence
\[ H = \left\{
\begin{bmatrix}
          1 & q \\
          0 & e
   \end{bmatrix}:
   \quad q\in\mathbb{H}
   \right\}, \quad
   S = \left\{
\begin{bmatrix}
          1 & 0 \\
          0 & s
   \end{bmatrix}:
   \quad s\in S^3
   \right\},
   \quad S\cap H = e
\leqno(4)\] 

First of all a subgroup $H$ is Abelian and isomorphic to $\mathbb{H}$ 
\[
   \begin{bmatrix}
          1 & q \\
          0 & e
   \end{bmatrix}
   \begin{bmatrix}
          1 & q' \\
          0 & e
   \end{bmatrix} =
   \begin{bmatrix}
          1 & q' + q e \\
          0 & e
   \end{bmatrix}
   =
   \begin{bmatrix}
          1 & q' + q \\
          0 & e
   \end{bmatrix}
\leqno(5)\]
and group $S$ is isomorphic to $S^3$
\[
   \begin{bmatrix}
          1 & 0 \\
          0 & s
   \end{bmatrix}
   \begin{bmatrix}
          1 & 0 \\
          0 & s'
   \end{bmatrix} =
   \begin{bmatrix}
          1 & 0 \\
          0 & s s'
   \end{bmatrix}
\leqno(6)\]

For the inner automorphism group we have
\[ I_g(g') = g g' g^{-1} =
   \begin{bmatrix}
          1 & (q' - q + q s')s^{-1} \\
          0 & s s' s^{-1}
   \end{bmatrix}
\leqno(7)\]

In particular
\[ H\ni h =
   \begin{bmatrix}
          1 & p \\
          0 & e
   \end{bmatrix}\longrightarrow
   I_g(h) =
   \begin{bmatrix}
          1 & p s^{-1} \\
          0 & e
   \end{bmatrix}\in H
\leqno(8)\]

I.e. $H$ --- normal divisor in $G$.

Any element of group $g\in G$ can be represented as \\
$g=s h,s\in S,h\in\mathbb{H}$.

Indeed
\[ g = s h =
   \begin{bmatrix}
          1 & 0 \\
          0 & s
   \end{bmatrix}
   \begin{bmatrix}
          1 & q \\
          0 & e
   \end{bmatrix} =
   \begin{bmatrix}
          1 & q  \\
          0 & s
   \end{bmatrix}
\leqno(9)\]
and the right side (9) completely defines $s,h$.

\bigskip
{\bf Remark 2}. We said that $G$ is a semidirect product \( G = S\circledS H\).
However the usual definition different from ours (see \cite[p. 119]{MarMisOrtPerRat07} and \cite[p. 119]{SWDiffGeom75}).
The usual semidirect product can be called as a {\it left} and this can be called as a {\it right}.

\newpage
\section{ The Lie algebra of group $G$ and adjoint representation }
\label{LieG}

The fact that group $G$ is a matrix group (albeit with non-commuting elements)
simplify find of adjoint representation of Lie bracket.

Obviously that Lie algebra $\bsym{g}$ of the group $G$ was formed of matrices
\[ \bsym{g} = \left\{ v =
   \begin{bmatrix}
          0 & \nu \\
          0 & \bsym{\xi}
   \end{bmatrix}:
   \quad \nu\in\mathbb{H},\quad \bsym{\xi}\in \mathbb{H}_0
   \right\}
\leqno(1)\]

The following notation will be considered equivalent
\[ v =
   \begin{bmatrix}
          0 & \nu \\
          0 & \bsym{\xi}
   \end{bmatrix} \simeq
   \begin{bmatrix}
         \nu\\
         \bsym{\xi}
   \end{bmatrix} \simeq
   (\nu,\bsym{\xi})\simeq \nu + \bsym{\xi}
\leqno(2)\]

Consider operator ${\rm Ad}$ of adjoint representation group $G$.

It can be found as differentiation with respect to argument $I_g(g')$ in (7)~\S3
when $q'=0,s'=e$ and also taking into account the matrix structure of group
by means of matrix automorphism applied to elements (1).
\[ {\rm Ad}_{(q,s)}[(\nu,\bsym{\xi})] =
   \begin{bmatrix}
          0 & (\nu + q\bsym{\xi})s^{-1}\\
          0 & s\bsym{\xi}s^{-1}
   \end{bmatrix} =
   \begin{bmatrix}
          0 & (\nu + q\bsym{\xi})s^\dag\\
          0 & s\bsym{\xi}s^\dag
   \end{bmatrix}
\leqno(3)\]

Operator ${\rm ad}$ now can be found as differentiation
with respect to $(q,s)$ when $(q,s)= (0,e)$
\[ {\rm ad}_{(\nu,\bsym{\xi})}
   \begin{bmatrix}
         \nu'\\
         \bsym{\xi}'
   \end{bmatrix} =
   \begin{bmatrix}
         \nu\bsym{\xi}' - \nu'\bsym{\xi}\\
         \bsym{\xi}\bsym{\xi}' - \bsym{\xi}' \bsym{\xi}
   \end{bmatrix} =
   \begin{bmatrix}
         \nu\bsym{\xi}' - \nu'\bsym{\xi}\\
         [\bsym{\xi}, \bsym{\xi}']
   \end{bmatrix}
\leqno(4)\]
that is consistent with (3a) \S2.

So we found the required Lie group $G$ .

\newpage
\section{ Co-adjoint representation group $G$ }
\label{CoAdjoint}

In the Lie algebra $\bsym{g}$ one can introduce a scalar product by the formula
\[  \langle v, v'\rangle
  = \frac12 {\rm tr}(v v' + v' v)
  = \frac12 (\nu\nu'^\dag + \nu^\dag\nu' + \bsym{\xi}\bsym{\xi}'^\dag + \bsym{\xi}^\dag\bsym{\xi}')
  = \langle\bsym{\xi},\bsym{\xi}'\rangle + \langle\nu,\nu'\rangle
\leqno(1)\]

{\bf Remark 3}. Scalar product of (1) allows us to identify vector spaces $\bsym{g}$ and $\bsym{g}^*$.

\bigskip
For quaternionic scalar product there are formulas
\[
\begin{cases}
   \langle a, q b\rangle =  \langle q^\dag a, b\rangle,\quad q,a,b\in\mathbb{H}; \\
   \langle a, b q\rangle =  \langle a q^\dag , b\rangle,\quad q,a,b\in\mathbb{H}\\
\end{cases}
\leqno(2)\]

Using definition (1) and properties (2) we find operator
adjoint to operator ${\rm Ad}$ in (3) \S4 with respect to scalar product (1).
\[ {\rm Ad}^*_{(q,s)^{-1}}
   \begin{bmatrix}
         \pi\\
         \bsym{\mu}
   \end{bmatrix} =
   \begin{bmatrix}
         \pi s^\dag\\
         s \left(\bsym{\mu} - \frac12(q^\dag\pi - \pi^\dag q)\right)s^\dag
   \end{bmatrix}
\leqno(3)\] 
or
\[ {\rm Ad}^*_{(q,s)^{-1}}
   \begin{bmatrix}
         \pi\\
         \bsym{\mu}
   \end{bmatrix} =
   \begin{bmatrix}
         \pi s^\dag\\
         s \left(\bsym{\mu} - q^\dag\pi\right)s^\dag + \langle q, \pi\rangle
   \end{bmatrix}
\leqno(3a)\]
where $(\pi,\bsym{\mu})\in \bsym{g}^*$ (see notations (2) \S4 and remark 3).

\newpage
\section{ Orbits of co-adjoint representation group~$G$ }
\label{Orbits}

Let's investigate orbits of co-adjoint representation group $G$.

Let $|\pi|=0\longrightarrow \pi=0$.

Then from formulas (3) or (3a) \S5 we have
\[ {\rm Ad}^*_{g^{-1}}
   \begin{bmatrix}
         0\\
         \bsym{\mu}
   \end{bmatrix} =
   \begin{bmatrix}
         0\\
         s \bsym{\mu}s^\dag
   \end{bmatrix} =
   \begin{bmatrix}
         0\\
         S[\bsym{\mu}]
   \end{bmatrix}
\leqno(1)\]
where $S$ --- an orthogonal transformation of 3-dimensional Euclidean space (rotation)
\cite{Zub_Quat1} and orbits diffeomorphic to spheres $S^2_{|\bsym{\mu}|}$.

\bigskip
Consider the basic case: $\rho=|\pi|\neq 0$.

Let's show that $\begin{bmatrix} |\pi|e_0\\ 0 \end{bmatrix}$
belongs to the orbit $\begin{bmatrix} \pi\\ \bsym{\mu} \end{bmatrix}$.

$\square$

Let's take
\[ s_0 = \frac{\pi}{|\pi|},\quad
  q_0 = -|\pi|^{-2}\pi\bsym{\mu}
\leqno(2)\]
then  $\pi s_0^\dag = |\pi|e_0$ and
\[ \langle q_0, \pi\rangle = -|\pi|^{-2}\langle\pi\bsym{\mu}, \pi\rangle
   = -|\pi|^{-2}\langle\bsym{\mu}, \pi^\dag \pi\rangle
   = -\langle\bsym{\mu}, e_0\rangle = 0
\leqno(3)\]
as well as
\[ q_0^\dag = |\pi|^{-2}\bsym{\mu}\pi^\dag\longrightarrow
   q_0^\dag\pi = \bsym{\mu}
\leqno(4)\]

From (3a) \S5 we get
\[ {\rm Ad}^*_{(q_0,s_0)^{-1}}
   \begin{bmatrix}
         \pi\\
         \bsym{\mu}
   \end{bmatrix} =
   \begin{bmatrix}
         |\pi|e_0\\
         0
   \end{bmatrix}
\leqno(5)\]

$\blacksquare$
\newpage
Using (1) and taking into account $\Im(q^\dag)=-\bsym{q}$
we obtain an arbitrary point of orbit in form
\[ {\rm Ad}^*_{(q,s)^{-1}}
   \begin{bmatrix}
         \rho e_0\\
         0
   \end{bmatrix} = \rho
   \begin{bmatrix}
      s^\dag\\
      s \bsym{q} s^\dag
   \end{bmatrix}
\leqno(6)\]

In particular this means that an arbitrary point of orbit we can obtained
by acting with subset of group elements $(\bsym{q}, s)$ on a fixed point $(\rho e_0, 0)$,
i.e. we use only pure quaternions.
\[ {\rm Ad}^*_{(\bsym{q}, s)^{-1}}
   \begin{bmatrix}
         \rho e_0\\
         0
   \end{bmatrix} = \rho
   \begin{bmatrix}
      s^\dag\\
      s \bsym{q} s^\dag
   \end{bmatrix}
\leqno(6a)\]

Thus, the above proved the following statement:

{\bf Proposition 1}. {\it
The orbits of co-adjoint representation group $G$ can be of two types:

1) Orbits of a point $(0, r \bsym{e}_3)$, $r \geq 0$
that are diffeomorphic to 2-dimensional spheres $S^2_r$,
$r$ --- a sphere radius;

2) Orbits $\mathcal{O}_\varrho$ of point $(\varrho \bsym{e}_0,0)$ , $\varrho > 0$,
that are diffeomorphic to $S^3_\varrho\times \mathbb{H}_0$,

where $S^3_\varrho$ --- a 3-dimensional sphere of radius $\varrho$.
}

\newpage
\section{ Symplectic structure on orbits of co-adjoint representation }
\label{Symplectic}

As for arbitrary Lie group on orbits of co-adjoint representation group $G$
exists symplectic 2-form of Kirillov-Kostant-Souriau (KKS).

For a fixed element of the Lie algebra $v=(\nu,\bsym{\xi})\in\bsym{g}$
at each point $(\pi,\bsym{\mu})$ of orbit the value of vector field
$v_{\bsym{g}^*}((\pi,\bsym{\mu}))$ determined that is infinitesimal generator
of co-adjoint group action $G$ (see \cite[p. 371]{Marsden:IMS94}).

Using (14.2.5) from \cite[p. 467]{Marsden:IMS94} and if differentiate (3a) \S5 we get
\[ v_{\bsym{g}^*}((\pi,\bsym{\mu}))
   = -{\rm ad}^*_{(\nu,\bsym{\xi})}
   \begin{bmatrix}
         \pi\\
         \bsym{\mu}
   \end{bmatrix} = -
   \begin{bmatrix}
         \pi \bsym{\xi}\\
         {\rm ad}^*_{\bsym{\xi}}[\bsym{\mu}]  + (\nu^\dag\pi - \langle\pi, \nu\rangle)
   \end{bmatrix}
\leqno(1)\]

Then symplectic 2-form KKS is determined from (14.3.1) \cite[p. 469]{Marsden:IMS94}
\[ \omega_{|(\pi,\bsym{\mu})}(v_{\bsym{g}^*}, v'_{\bsym{g}^*})
 = -\langle(\pi,\bsym{\mu}), [v, v']\rangle
\leqno(2)\]
\[ = -\langle(\pi,\bsym{\mu}), (\nu\bsym{\xi}' - \nu'\bsym{\xi}, [\bsym{\xi}, \bsym{\xi}'])\rangle
   = -\langle \pi, \nu\bsym{\xi}' - \nu'\bsym{\xi}\rangle
   -\langle\bsym{\mu}, [\bsym{\xi}, \bsym{\xi}']\rangle
\]

By definition an arbitrary symplectic 2-form is closed.
By Darboux's theorem each that sort of 2-form is a locally exact,
i.e. is an exterior differential of 1-form.

However, neither in general nor in KKS case these forms are not must to be a globally accurate.
Nevertheless for orbits of 2-nd type (i.e., $|\pi|\neq 0$) form (2) is exact.

Namely: let's consider the following expression for 1-form $\theta$
\[ \theta_{|(\pi,\bsym{\mu})}(v_{\bsym{g}^*}((\pi,\bsym{\mu})))
   = -\langle\bsym{\mu},\bsym{\xi}\rangle, \quad
   v = (\nu,\bsym{\xi})
\leqno(3)\]

Form $\theta$ where $|\pi|\neq 0$ is defined correct because from (1) we can see
\[ v_{\bsym{g}^*}=v'_{\bsym{g}^*}\longrightarrow\bsym{\xi}~=~\bsym{\xi}'
   \longrightarrow\langle\bsym{\mu},\bsym{\xi}\rangle = \langle\bsym{\mu},\bsym{\xi}'\rangle
\]

Let's use well known formula for exterior differential of 1-form
\[ d\theta_{|(\pi,\bsym{\mu})}(v_{\bsym{g}^*}, v'_{\bsym{g}^*})
   = \partial_{v_{\bsym{g}^*}}\theta(v'_{\bsym{g}^*})
   - \partial_{v'_{\bsym{g}^*}}\theta(v_{\bsym{g}^*})
   - \theta([v_{\bsym{g}^*}, v'_{\bsym{g}^*}])
\leqno(4)\]

For infinitesimal generators the formula is fulfilled (11.1.5) \cite[p. 372]{Marsden:IMS94}
\[ [v_{\bsym{g}^*}, v'_{\bsym{g}^*}] = -[v, v']_{\bsym{g}^*}
\leqno(5)\]

Taking into account (5) and the fact
that scalar and pure quaternions are orthogonal
then from (3) and (4) we have
\[ d\theta_{|(\pi,\bsym{\mu})}(v_{\bsym{g}^*}, v'_{\bsym{g}^*})
   = -\partial_{v_{\bsym{g}^*}}\langle\bsym{\mu},\bsym{\xi}'\rangle
   + \partial_{v'_{\bsym{g}^*}}\langle\bsym{\mu},\bsym{\xi}\rangle
   - \langle\bsym{\mu},[\bsym{\xi}, \bsym{\xi}']\rangle
\]
\[ = \langle{\rm ad}^*_{\bsym{\xi}}[\bsym{\mu}]  + (\nu^\dag\pi - \langle\pi, \nu\rangle),\bsym{\xi}'\rangle
   - \langle{\rm ad}^*_{\bsym{\xi}'}[\bsym{\mu}]  + (\nu'^\dag\pi - \langle\pi, \nu'\rangle),\bsym{\xi}\rangle
   - \langle\bsym{\mu},[\bsym{\xi}, \bsym{\xi}']\rangle
\]
\[ = \langle{\rm ad}^*_{\bsym{\xi}}[\bsym{\mu}]  + \nu^\dag\pi,\bsym{\xi}'\rangle
   - \langle{\rm ad}^*_{\bsym{\xi}'}[\bsym{\mu}]  + \nu'^\dag\pi,\bsym{\xi}\rangle
   - \langle\bsym{\mu},[\bsym{\xi}, \bsym{\xi}']\rangle
\]
\[ = \langle{\rm ad}^*_{\bsym{\xi}}[\bsym{\mu}],\bsym{\xi}'\rangle
   + \langle\nu^\dag\pi,\bsym{\xi}'\rangle
   - \langle{\rm ad}^*_{\bsym{\xi}'}[\bsym{\mu}],\bsym{\xi}\rangle
   - \langle\nu'^\dag\pi,\bsym{\xi}\rangle
   - \langle\bsym{\mu},[\bsym{\xi}, \bsym{\xi}']\rangle
\]
\[ = \langle{\bsym{\mu},\rm ad}_{\bsym{\xi}}[\bsym{\xi}']\rangle
   + \langle\pi,\nu\bsym{\xi}'\rangle
   - \langle\bsym{\mu},{\rm ad}_{\bsym{\xi}'}[\bsym{\xi}\rangle]
   - \langle\pi,\nu'\bsym{\xi}\rangle
   - \langle\bsym{\mu},[\bsym{\xi}, \bsym{\xi}']\rangle
\]
\[ = \langle\bsym{\mu},[\bsym{\xi}, \bsym{\xi}']\rangle
   - \langle\pi,\nu'\bsym{\xi}\rangle
   + \langle\pi,\nu\bsym{\xi}'\rangle
\]
i.e.
\[ -d\theta_{|(\pi,\bsym{\mu})}(v_{\bsym{g}^*}, v'_{\bsym{g}^*})
   = -\langle(\pi, \bsym{\mu}),(\nu\bsym{\xi}' + \nu'\bsym{\xi}, [\bsym{\xi}, \bsym{\xi}'])\rangle
   = \omega_{|(\pi,\bsym{\mu})}(v_{\bsym{g}^*}, v'_{\bsym{g}^*})
\leqno(6)\]

Thus above we proved the following statement:

{\bf Proposition 2}. {\it
On the orbits of co-adjoint representation group $G$ of 2-nd type (i.e. $|\pi|\neq 0$) 2-form KKS is exact,
i.e. there exists a form $\theta$ such that $\omega = -d\theta$,
where $\omega$ and $\theta$ are given by expressions (2) and (3).
}

\newpage
\section{ The orbit of 2-nd type of co-adjoint representation group $G$ is symplectically isomorphic to $T^*(S^3_{\varrho})$ }
\label{Sphere}

Proposition 1 establishes diffeomorphism of 2-nd type orbit of co-adjoint representation group $G$ to direct product of $S^3_\varrho\times \mathbb{R}^3$.

As it shown in Part I cotangent bundle $T^*(S^3)$ ($S^3=S^3_1$) has form of direct product $S^3\times \mathbb{H}_0$
in representation of left (right) trivialization.

Sphere $S^3_\varrho$ in contrast to $S^3$ not a group with respect to multiplication of quaternions,
but trivialization of $T^*(S^3_{\varrho})$ can be performed in a similar way.

\bigskip
Namely (see (3) of previous section):
\[  \langle\bsym{\mu},\bsym{\xi}\rangle
    = \left\langle\frac{\pi\bsym{\mu}}{|\pi|^2},\pi\bsym{\xi}\right\rangle
    = \frac{1}{\varrho^2}\left\langle\pi\bsym{\mu},\pi\bsym{\xi}\right\rangle
\leqno(1)\]

Vector $\pi\bsym{\xi}$ is an arbitrary tangent to sphere $S^3_{\varrho}$
therefore quaternion $\frac{\pi\bsym{\mu}}{|\pi|^2}$ can be interpreted as an element of cotangent bundle $T^*(S^3_{\varrho})$.

Thus there is a diffeomorphism
\[ \phi: \mathcal{O}_\varrho\ni -(\pi, \bsym{\mu})\rightarrow \alpha_\pi \in T^*(S^3_{\varrho}):
   \quad \alpha_\pi = -\frac{1}{\varrho^2}\pi\bsym{\mu}
\leqno(2)\]

It is obvious that for so-defined diffeomorphism $\phi$ we have
\[ \theta = \phi^*\Theta
\leqno(3)\]

Thus we have the following statement:

{\bf Proposition 3}. {\it
Symplectic structure on 2-nd type orbit $\mathcal{O}_\varrho$ of co-adjoint representation group $G$
is symplectically diffeomorphic to canonical symplectic structure on cotangent bundle $T^*(S^3_{\varrho})$.
}

\newpage
It is necessary to make a few comments
that clarifies the meaning of derived above equivalence of different by construction Poisson structures.

Initially, quaternionic Poisson structures in (1) \S2 were obtained in the frame of study the Lie-Poisson structure of group $SE(4)$ in \cite{BM:RBD01}, where they are one subsets of Poisson structures in this structure.

In this work we found Lie group and Lie-Poisson structure that is strictly correspond to the Poisson brackets in (1) \S2.

In Part I the same Poisson brackets were received
as canonical Poisson brackets in Hamiltonian mechanics of a rigid body.

Quaternion variables $q$ in (1) \S2 have sense of group parameters that define the position of a rigid body (Rodrigues-Hamilton formula) at that the parameters have restriction $|q|=1\longrightarrow q\in S^3_1=S^3$.

Proposition 3 establishes a mathematical equivalence of the Lie-Poisson structure and canonical Poisson structure on $T^*(S^3)$.
In process the initial group quaternion variables $q$ in the Lie-Poisson structure are replaced on quaternion variables $\pi\in\bsym{g}^*$.
There is no restriction on $|\pi|=1$ if we go from idea of the approach.
Variables $q$ and $\pi$ therefore have different geometrical and mechanical sense.

Thus direct mechanical meaning for a rigid body dynamics has the canonical Poisson structure was obtained in Part I
whereas Lie-Poisson structures that was obtained in work \cite{BM:RBD01}
and here are alternative approaches to dynamics of a rigid body
that gives more opportunities to use group-theoretic or more general algebraic methods of investigation.

\end{document}